\documentclass{article}
\usepackage{spconf,amsmath,graphicx,hyperref}
\usepackage{cite}
\usepackage{algorithmic}
\usepackage{amssymb}
\usepackage{graphicx}
\usepackage{textcomp}
\usepackage{xcolor}
\usepackage{booktabs}
\usepackage{makecell}
\usepackage{color}
\usepackage{colortbl}
\usepackage{multirow}
\usepackage{multicol}
\usepackage{adjustbox}
\usepackage{arydshln}
\usepackage[flushleft]{threeparttable}
\usepackage{hyperref}
\usepackage{xcolor}
\usepackage{subcaption}
\definecolor{myblue}{RGB}{108, 142, 191}
\hypersetup{
    colorlinks=true,
    linkcolor=myblue,
    citecolor=myblue,
    filecolor=myblue,
    urlcolor=myblue
}

\title{Localizing Speech Deepfakes Beyond Transitions via\\ Segment-Aware Learning}

\name{Yuchen Mao$^1$ \qquad Wen Huang$^1$ \qquad Yanmin Qian$^{1,2}$
\sthanks{This work was supported in part by China NSFC project under Grants No. U25A20409, and in part by SJTU Med-X (Medicine \& Engineering) Translational Research Grant (YG2025LC09).}
\sthanks{ Yanmin Qian is the corresponding author.}}
\address{$^1$Auditory Cognition and Computational Acoustics Lab\\
MoE Key Lab of Artificial Intelligence, AI Institute\\
School of Computer Science, Shanghai Jiao Tong University, Shanghai, China \\
$^2$VUI Labs
}

\begin{document}
\ninept
\maketitle
\begin{abstract}

Localizing partial deepfake audio, where only segments of speech are manipulated, remains challenging due to the subtle and scattered nature of these modifications.
Existing approaches typically rely on frame-level predictions to identify spoofed segments, and some recent methods improve performance by concentrating on the transitions between real and fake audio.
However, we observe that these models tend to over-rely on boundary artifacts while neglecting the manipulated content that follows.
We argue that effective localization requires understanding the entire segments beyond just detecting transitions. Thus, we propose Segment-Aware Learning (SAL), a framework that encourages models to focus on the internal structure of segments. 
SAL introduces two core techniques: Segment Positional Labeling, which provides fine-grained frame supervision based on relative position within a segment; and Cross-Segment Mixing, a data augmentation method that generates diverse segment patterns.
Experiments across multiple deepfake localization datasets show that SAL consistently achieves strong performance in both in-domain and out-of-domain settings, with notable gains in non-boundary regions and reduced reliance on transition artifacts.
The code is available at \url{https://github.com/SentryMao/SAL}.


\end{abstract}
\begin{keywords}
partial spoof, speech deepfake localization, anti-spoofing, segment-aware learning
\end{keywords}
\vspace{-1mm}
\section{Introduction}
\label{sec:intro}

Recent advancements in speech synthesis and voice conversion have enabled the creation of highly realistic speech deepfakes, in which audio is artificially generated or manipulated to resemble genuine human speech. While current countermeasures have demonstrated strong performance in detecting fully spoofed utterances\cite{wang2021comparative,yi2023audio,ge2025post,huang2025data}, a more subtle and pressing threat is partial spoofing\cite{zhang2022partialspoof,yi2021half,yi2022add,yi2023add}. In this scenario, only specific segments of an utterance are manipulated, such as altering critical words in a recorded statement. These fine-grained edits are harder to detect and often more dangerous in real-world scenarios. To defend against such attacks, it is essential not only to identify whether an utterance is fake, but to precisely localize the manipulated regions within the audio stream.

Prior approaches to this task primarily employed frame-level detection, where a model renders a binary prediction for each acoustic frame based on localized acoustic cues\cite{zhang2021initial,zhang2021multi,yadav2024mdrt,zhang2022partialspoof}. More recently, transition-aware methods have achieved superior results by focusing on transition regions, which are the boundaries between real and fake segments and often contain subtle artifacts introduced during the splicing and overlap operation in the dataset construction. For instance, BAM\cite{zhong2024enhancing} employs a boundary enhancement module and frame-wise attention to emphasize features near transition points. AGO\cite{zeng2025adversarial} jointly optimizes region classification and boundary detection using a shared backbone and adversarial gradient updates. CFPRF\cite{wu2024coarse} enhances boundary features by modeling context across multiple transition points through a dedicated enhancement module. BFC-Net\cite{zhou2025bfc} integrates multi-scale features and applies boundary-frame cross-attention to improve prediction consistency around segment edges.


A key finding from Liu et al.\cite{liu2024neural} offers an explanation for the success of transition-aware learning. Using Grad-CAM\cite{selvaraju2017grad} and a quantitative analysis metric, they show that many strong-performing countermeasures consistently focus on transition regions as the primary artifacts for detection. Notably, they observe that models tend to overlook these regions when making incorrect predictions, whereas correct predictions are characterized by a stronger focus on these areas. This underscores the critical role of transition regions in accurately localizing partially spoofed audio.


\begin{figure}
    \centering
    \includegraphics[width=\linewidth]{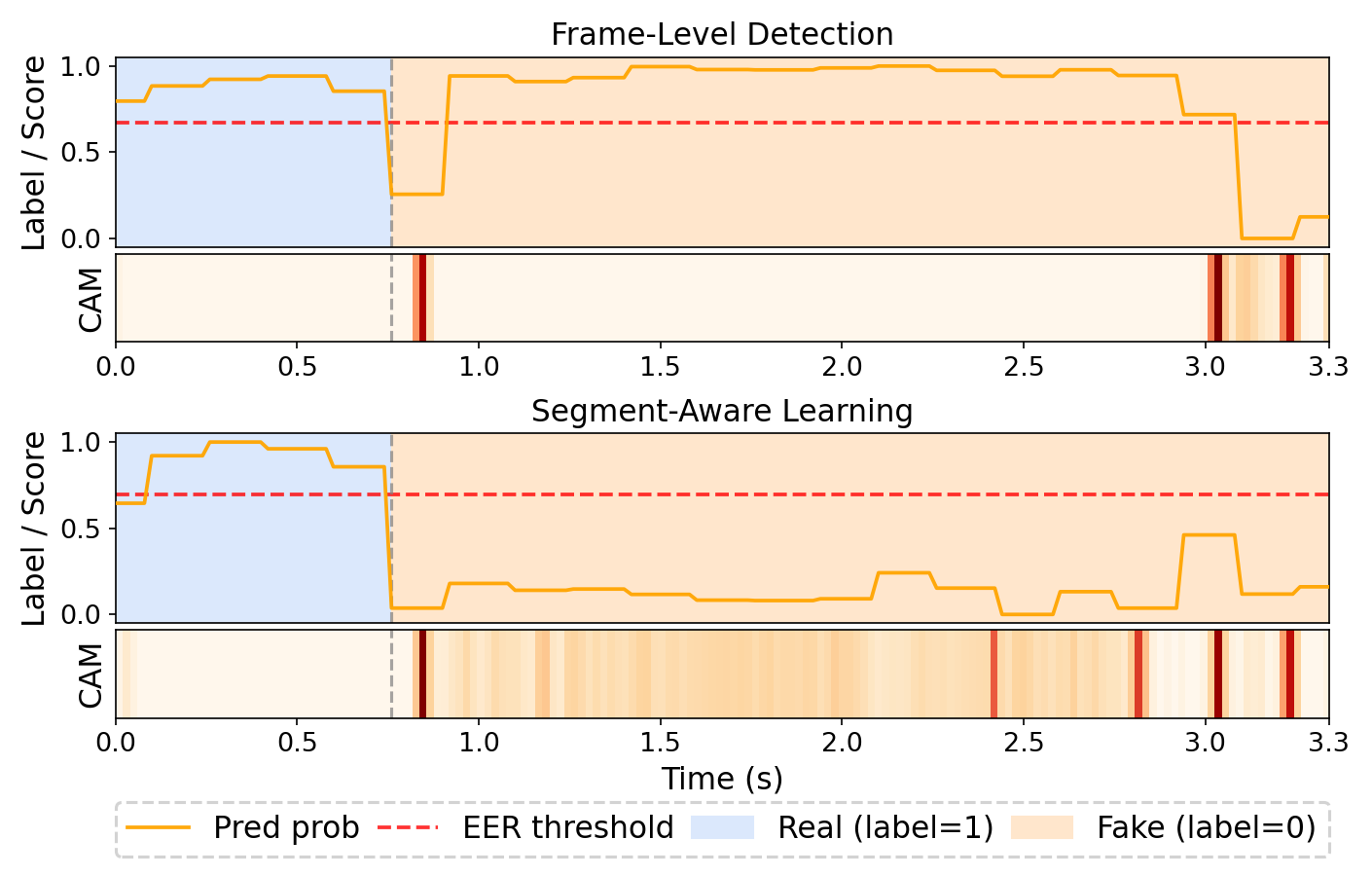}  
    \caption{Prediction scores and Grad-CAM visualizations (deeper color indicates higher values) for \texttt{CON\_E\_0000096.wav} in PartialSpoof evaluation set, comparing the baseline frame-level detection system (top) and our segment-aware learning system (bottom).}
    \label{fig:intro}
\vspace{-20pt}
\end{figure}

While the importance of transition regions should not be overlooked, an over-reliance on them may introduce unintended side effects. Specifically, we observe that models often focus excessively on boundary areas, which leads to a form of shortcut learning that undermines generalization. In our experiments with a strong frame-level detection baseline, we observed that many misclassified cases share a common pattern: the model correctly detects transitions but fails to recognize the manipulated content that follows. Fig.~\ref{fig:intro} shows a typical example. In the Grad-CAM visualization, the baseline model’s attention is concentrated almost exclusively in the form of sharp spikes at the transition boundaries of the spoofed segment, with negligible focus on the content within the segment itself. This behavior explains the poor prediction results in the middle part, where a large continuous region is misclassified as genuine. These findings motivate a shift away from transition-centric learning to a more comprehensive strategy: instead of relying solely on transition artifacts, the model is encouraged to capture meaningful features across entire segments for accurate deepfake localization.



In this paper, we propose \textbf{Segment-Aware Learning (SAL)}, a  framework designed to move beyond the detection of transition artifacts and instead model the intrinsic characteristics of entire audio segments. SAL introduces two primary techniques: Segment Positional Labeling (SPL), a novel fine-grained labeling scheme that supervises each frame with both its binary class and its relative position within a segment; and Cross-Segment Mixing (CSM), a data augmentation technique that creates more diverse segment patterns by cutting and mixing segments from different utterances. 
Through experiments, the proposed SAL framework demonstrates consistently strong performance across multiple deepfake localization datasets. Furthermore, it effectively reduces shortcut learning and significantly improves the localization of spoofed segments, particularly in non-boundary regions that are often overlooked.





\vspace{-1mm}
\section{Segment-Aware Learning}
\label{sec:intro}

\begin{figure}
    \centering
    \includegraphics[width=0.9\linewidth]{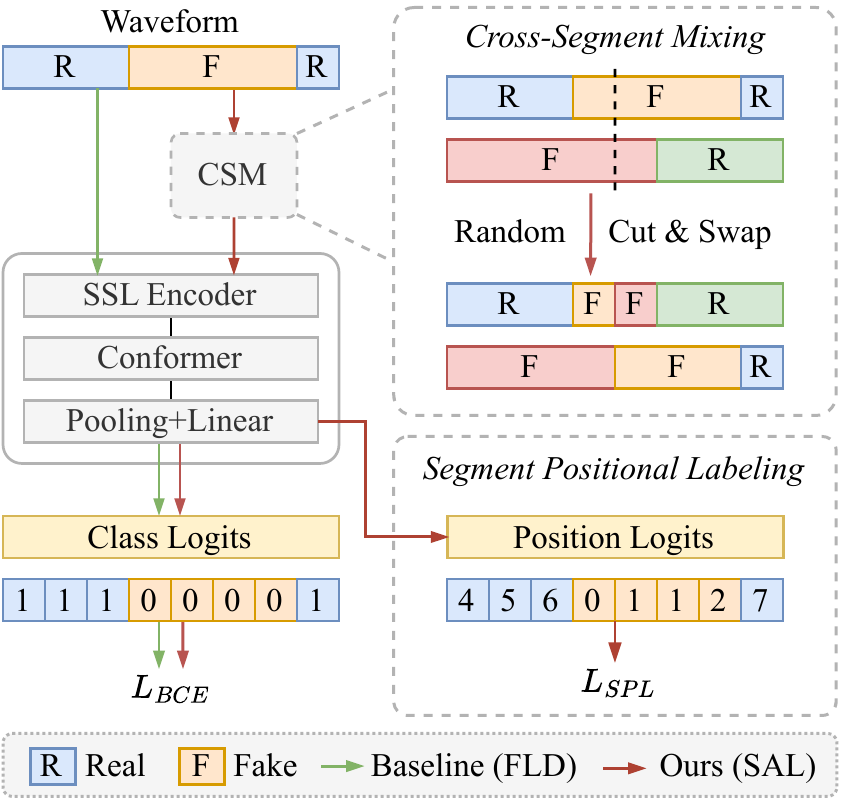}   
    \caption{Overview of the proposed Segment-Aware Learning (SAL) framework. Arrows in different colors indicate the distinction between the baseline Frame-Level Detection (FLD) pipeline and the proposed SAL strategy.}
    \label{fig:sal}
\vspace{-17pt}
\end{figure}

Segment-Aware Learning is designed to shift the model's focus from solely identifying boundary artifacts to understanding the internal features of audio segments. This framework enhances a standard frame-level classification backbone with two techniques: Segment Positional Labeling, a multi-task supervision strategy, and Cross-Segment Mixing, a data augmentation technique. Fig.~\ref{fig:sal} illustrates the overview of the framework.

\vspace{-2mm}
\subsection{Segment Positional Labeling}

Traditional partial spoof detection models are trained using simple binary frame-level labels (e.g., $1$ for real, $0$ for fake). This sparse supervision signal is insufficient to enforce segment-level coherence. 

To counteract this, we introduce SPL as an auxiliary supervision. It reframes the task as a multi-label problem, forcing the model to learn not just what a frame is real or fake but also where it is within its local context. To achieve this, we categorize each frame based on its relative position within a continuous segment of the same class using four positional labels: Start ($S$), Middle ($M$), End ($E$), and Unit ($U$) for single-frame segments. By combining the two binary class labels, Real($R$) or Fake($F$), with these four positional labels, this approach creates a total of eight distinct label combinations for any given frame, such as $(F, S)$ or $(R, M)$. This mechanism is implemented using an auxiliary prediction head in the model.

The model is trained with a combined loss that balances two objectives: standard binary classification and segment positional labeling.  The total loss $L_{\text{total}}$ is a weighted sum of the Binary Cross-Entropy loss ($L_{\text{BCE}}$) for the real / fake prediction and a Cross-Entropy loss ($L_{\text{SPL}}$) for the positional label prediction:
\begin{equation}
    L_{\text{total}} = L_{\text{BCE}} + \lambda L_{\text{SPL}}
    \label{eq:total_loss}
\end{equation}
where $\lambda$ is a hyperparameter that balances the contribution of the two loss terms.

By forcing the model to explicitly predict positional context, SPL discourages over-reliance on boundary cues. The model learns the stable, representative characteristics of a continuous segment, thereby improving its understanding of local temporal dynamics and acoustic coherence.

\vspace{-2mm}
\subsection{Cross-Segment Mixing}

Training datasets for partial spoof detection often contain inherent biases in length, position, and type of spoofed segments. A model trained on such data may fail to generalize to unseen segment compositions. CSM is a data augmentation strategy designed to mitigate this overfitting and disrupt the model's reliance on recurring transition patterns.

As shown in Fig.~\ref{fig:sal}, CSM generates novel training samples by splicing together two audio utterances. Specifically, let $(x_A, y_A)$ and $(x_B, y_B)$ be two training samples randomly selected from a mini-batch. Here, $x_A, x_B \in \mathbb{R}^T$ are one-dimensional audio sequences of length $T$, and $y_A, y_B \in \mathbb{R}^{T \times K}$ are their corresponding frame-level label sequences. The objective is to generate a new hybrid training sample $(\tilde{x}, \tilde{y})$ by concatenating both the input sequence and the label sequence at the same crossover point. This operation is defined as:
\begin{align}
    \tilde{x} &= \mathbf{M} \odot x_A + (\mathbf{1} - \mathbf{M}) \odot x_B \\
    \tilde{y} &= \mathbf{M} \odot y_A + (\mathbf{1} - \mathbf{M}) \odot y_B
\end{align}
where $\odot$ denotes element-wise multiplication. The vector $\mathbf{M} \in \{0, 1\}^T$ is a shared binary mask that determines which parts of the sequences are combined. It is generated by first sampling a random integer crossover point $\lambda_c \sim \text{Uniform}\{1, \dots, T-1\}$. Here, $\mathbf{1}$ is a vector of ones of length $T$. The mask $\mathbf{M}$ is then defined for each time step $t$ as:
\begin{align}
\mathbf{M}(t) =
\begin{cases}
1 & \text{if } t \le \lambda_c \\
0 & \text{if } t > \lambda_c
\end{cases}
\end{align}

The frame-level labels of the newly generated sample, including binary class and SPL positional labels, are updated by combining corresponding labels from the original utterances. At the new splice point, positional labels are re-assigned to reflect their updated roles, while the internal labels of the pasted segment remain unchanged. Importantly, if the two segments at the splice point share the same class, the transition frame is labeled as `Middle' rather than being treated as a boundary. This prevents the model from overemphasizing intra-class transitions and helps it to learn consistent representations within continuous segments.

During training, the mixing process can be applied multiple times using different pairs of sampled utterances to create finer-grained segment variations. In practice, we set a fixed mixing probability and a maximum number of mixing rounds to generate augmented samples while retaining the original ones.

CSM exposes the model to a much wider variety of segment lengths, positions, and boundary types than are present in the original dataset. This diverse exposure disrupts shortcut learning by preventing the model from memorizing specific boundary patterns. Simultaneously, it augments the training data with challenging and less frequent patterns, such as very short single-frame spoofs or extremely long continuous segments, ultimately leading to a more resilient and generalizable countermeasure.


\vspace{-1mm}
\section{Experiments and Analysis}
\label{sec:intro}
\subsection{Experimental Settings}

\textbf{Datasets:} Three distinct datasets are used in our experiments: PartialSpoof (PS)\cite{zhang2022partialspoof}, the Half-truth Audio Detection (HAD) dataset\cite{yi2021half} and LlamaPartialSpoof (LPS) dataset\cite{luong2025llamapartialspoof}. For PS and HAD, we train and evaluate our models on their respective official splits. To assess robustness against unseen scenarios, we also perform a cross-dataset evaluation on the LPS (crossfade version), using the model trained solely on PS.

\noindent\textbf{Model Architecture:} The core architecture of our SAL framework uses a pre-trained SSL model as the front-end feature extractor. We experiment with two front-ends: Wav2Vec2-XLSR\cite{babu2021xls} and WavLM-Large\cite{chen2022wavlm}. For ablation studies, we use the Wav2Vec2-XLSR model. The features from the SSL model are optionally passed through a lightweight Conformer module\cite{gulati2020conformer} (2 blocks, 4 attention heads per block). To align with the target resolution, we apply average pooling to downsample the original 20 ms frame-level features, followed by MLP layers for final predictions.

\noindent\textbf{Training Configuration:} Our input consists of raw audio waveforms sampled at 16 kHz. During training, the resolution for the PS dataset is 160 ms, while for HAD it is 20 ms. All audio segments are padded or truncated to a fixed length of 4 seconds. To enhance model robustness, we apply RawBoost\cite{tak2022rawboost} to the input with a probability of 0.5 and CSM with a probability of 0.2 during training. All models are trained for up to 50 epochs using the Adam optimizer with a learning rate of $1 \times 10^{-5}$, a weight decay of 0.0001, and a batch size of 32. A StepLR scheduler is employed to adjust the learning rate, reducing it by a factor of 0.1 every 10 epochs. The loss-balancing hyperparameter $\lambda$ for the SPL component is set to 0.1.

\noindent\textbf{Evaluation Configuration:} We evaluate system performance using two primary metrics: the frame-level Equal Error Rate (EER) and the F1-score. The model checkpoint that achieves the lowest EER on the development set is selected for final testing. To ensure a fair comparison with prior work, the test resolution is set to 160 ms for the PS dataset, and 20 ms for both the HAD and LPS datasets.

\subsection{Performance Comparison}

To validate the effectiveness of our proposed SAL framework, we compare its performance with existing methods on three distinct datasets: PS, HAD, and LPS.

\begin{table}[htbp]
\centering
\caption{Metrics (\%) of different methods trained and evaluated on PS. Results for other methods are derived from the original papers.}
\label{tab:ps}
\vspace{-8pt}
\resizebox{0.95\linewidth}{!}{%
\begin{tabular}{l|l|cc}
\Xhline{1pt}
\textbf{System} & \textbf{Front-end} & \textbf{EER$\downarrow$} & \textbf{F1$\uparrow$} \\
\hline
LCNN-BLSTM\cite{zhang2021initial}      & LFCC              & 16.21     & -      \\
SELCNN-BLSTM\cite{zhang2021multi}    & LFCC              & 15.93     & -      \\
GNCL\cite{ge2025gncl}            & W2V2-Base         & 11.81     & -      \\
MDRT\cite{yadav2024mdrt}            & W2V2-Base+M2D     & 8.82      & -      \\
Multi reso.\cite{zhang2022partialspoof}  & W2V2-Large        & 9.24      & -      \\
TDL\cite{xie2024efficient}             & W2V2-XLSR         & 7.04      & 88.96  \\
SPF\cite{cai2024integrating}             & WavLM             & -         & 92.96  \\
CFPRF\cite{wu2024coarse}           & W2V2-XLSR         & 7.41      & 93.89  \\
AGO\cite{zeng2025adversarial}             & W2V2-XLSR         & 6.79      & 94.36  \\
BAM\cite{zhong2024enhancing}             & W2V2-XLSR         & 4.12      & 94.98  \\
BAM\cite{zhong2024enhancing}             & WavLM             & 3.58      & 96.09  \\
BFC-Net\cite{zhou2025bfc}         & W2V2-XLSR         & 3.41      & -      \\
BFC-Net\cite{zhou2025bfc}         & WavLM             & \textbf{2.73}      & 96.69  \\
\hline
SAL            & W2V2-XLSR         & 3.32      & \underline{96.84}  \\
SAL            & WavLM             & \underline{3.00}         & \textbf{97.09}      \\
\Xhline{1pt}
\end{tabular}}
\end{table}


As presented in Table~\ref{tab:ps}, our proposed SAL framework demonstrates highly competitive performance on the PartialSpoof benchmark. Notably, our method achieves the highest F1-score among all listed systems, showcasing a superior balance of precision and recall. While its EER is comparable to other top-performing solutions, SAL consistently outperforms alternative boundary-focused systems such as BAM and AGO.

\begin{table}[ht]
\centering
\caption{Metrics (\%) of different methods trained and evaluated on HAD. Results for other methods are derived from \cite{wu2024coarse}.}
\vspace{-8pt}
\label{tab:had}
\resizebox{0.8\linewidth}{!}{%
\begin{tabular}{l|l|cc}
\Xhline{1pt}
\textbf{System} & \textbf{Front-end} & \textbf{EER$\downarrow$} & \textbf{F1$\uparrow$} \\
\hline
SPF\cite{cai2024integrating}         & WavLM        & 0.35         & 99.78 \\
Multi reso.\cite{zhang2022partialspoof} & W2V2-Large   & 0.18         & 99.89 \\
CFPRF\cite{wu2024coarse}               & W2V2-XLSR    & 0.08         & 99.95 \\
\hline
SAL                & W2V2-XLSR    & \textbf{0.05}         & \textbf{99.99} \\
SAL                & WavLM        & \underline{0.05}      & \underline{99.99}      \\
\Xhline{1pt}
\end{tabular}}
\end{table}


Further validation on the HAD dataset confirms the superiority of our framework (Table~\ref{tab:had}). The proposed SAL model establishes a new state-of-the-art, decisively outperforming all prior systems on both EER and F1-score. This robust performance is consistently achieved with both W2V2-XLSR and WavLM front-ends, underscoring the model's strong generalization capabilities across different acoustic feature extractors.


\begin{table}[ht]
\centering
\caption{Metrics (\%) of different methods trained on PS and evaluated on LPS.  Results for Multi reso. are reported by \cite{luong2025llamapartialspoof}. Results for BAM are evaluated using the released checkpoint.}
\label{tab:lps}
\vspace{-8pt}
\resizebox{0.85\linewidth}{!}{%
\begin{tabular}{l|l|cc}
\Xhline{1pt}
\textbf{System} & \textbf{Front-end} & \textbf{EER$\downarrow$} & \textbf{F1$\uparrow$} \\
\hline
Multi reso.\cite{zhang2022partialspoof, luong2025llamapartialspoof} & W2V2-Large   & 47.49         & - \\
BAM\cite{zhong2024enhancing}                & WavLM    & 42.58         & 53.40 \\
\hline
SAL                & W2V2-XLSR    &  \textbf{35.52}    & \underline{55.30} \\
SAL                & WavLM        &  \underline{36.60}    & \textbf{56.09}      \\
\Xhline{1pt}
\end{tabular}}
\end{table}


To assess robustness against unseen and more challenging scenarios, we evaluated our model on the LlamaPartialSpoof dataset, with results in Table~\ref{tab:lps}. While all systems exhibited performance degradation on this difficult out-of-distribution task, our SAL framework significantly outperformed all baselines. By achieving the best performance in both EER and F1-score, our model demonstrates superior generalization.

\vspace{-2mm}

\subsection{Ablation Study}

\vspace{-1mm}

We conduct a comprehensive ablation study to dissect the contribution of each key component in our proposed architecture. The analysis is performed on both the PS and the more challenging LPS datasets, with detailed results presented in Table~\ref{tab:ablation}.



\begin{table}[ht]
\centering
\vspace{-2mm}
\caption{Ablation study on architecture and training strategies for systems trained on PS and evaluated on PS and LPS (metrics in \%).}
\label{tab:ablation}
\vspace{-8pt}
\resizebox{\linewidth}{!}{%
\begin{tabular}{c:l|cc|cc}
\Xhline{1pt}
\multicolumn{2}{l|}{\multirow{2}{*}{\textbf{System Configuration}}} & \multicolumn{2}{c|}{\textbf{PS}} & \multicolumn{2}{c}{\textbf{LPS}} \\
\cline{3-6}
\multicolumn{2}{c|}{} & \textbf{EER$\downarrow$} & \textbf{F1$\uparrow$} & \textbf{EER$\downarrow$} & \textbf{F1$\uparrow$} \\
\hline
\rowcolor{gray!10}
\multicolumn{6}{l}{\textbf{\textit{Base Model Enhancements}}} \\
\textit{S0} & SSL + Avg Pool & 5.70 & 94.35 & 42.24 & 42.69 \\
\textit{S1} & \textit{S0} + Layer Weighting & 5.28 & 94.69 & 41.57 & 45.29 \\
\textit{S2} & \textit{S1} + Conformer & 5.05 & 94.65 & 42.33 & 46.32 \\
\textit{S3} & \textit{S2} + RawBoost & 3.95 & 96.23 & 38.62 & 51.81 \\
\rowcolor{gray!10}
\multicolumn{6}{l}{\textbf{\textit{Segment Positional Labeling}}} \\
\textit{S4} & \textit{S3} + Transition Loss & 3.81 & 96.33 & 36.11 & 55.03 \\
\textit{S5} & \textit{S3} + Position Loss & 3.50 & 96.62 & \textbf{35.33} & \textbf{55.58} \\
\rowcolor{gray!10}
\multicolumn{6}{l}{\textbf{\textit{Cross-Segment Mixing}}} \\
\textit{S6} & \textit{S5} + CSM (1 round) & \underline{3.42} & \underline{96.69} & 36.17 & 54.63 \\
\textit{S7} & \textit{S5} + CSM (2 rounds) & \textbf{3.32} & \textbf{96.84} & 35.52 & \underline{55.30} \\
\textit{S8} & \textit{S5} + CSM (3 rounds) & 3.57 & 96.58 & \underline{35.48} & 55.08 \\
\Xhline{1pt}
\end{tabular}%
}
\vspace{-5pt}
\end{table}

Our investigation begins with a simple baseline (S0) consisting of an SSL front-end and an average pooling layer. We observe steady performance gains by incrementally incorporating weighted layer averaging (S1) and a Conformer block (S2). The most substantial improvement in this stage comes from applying RawBoost data augmentation (S3), which drastically reduces the EER on both datasets.

Building upon the enhanced baseline (S3), we evaluate our core SAL strategies. The introduction of the Transition Loss (S4), which models boundary segments, further improves performance. However, applying the Position Loss (S5) yields a more significant gain, achieving the best overall results on the challenging LPS dataset. This finding underscores the importance of explicitly encoding the absolute position of each segment for robust spoof detection, especially in out-of-distribution scenarios.

Finally, we analyze the impact of our Cross-Segment Mixing (CSM) module, which is designed to capture non-local segment interactions. When applied to the position-loss-based system (S5), even a single round of CSM (S6) provides a boost. Increasing the complexity to two rounds (S7) hits a sweet spot, achieving the best overall performance on the PS dataset and a highly competitive result on LPS. Adding a third round (S8) slightly degrades performance, suggesting that two rounds of mixing strike an optimal balance between enhancing data diversity and preserving feature integrity.This validates the architecture of our final proposed model (S7).
\vspace{-2mm}
\subsection{Visualization and Analysis}
\vspace{-1mm}

To reveal the underlying mechanism of our model's decision-making process, we begin with an analysis of a representative case. Fig.~\ref{fig:intro} uses Grad-CAM to visualize the difference in attention between a traditional FLD model and our SAL model. The frame-level attributions were generated using Grad-CAM, for which the target score was calculated by summing the fake probabilities across all frames of the utterance. This approach helps to visualize which frames contribute most to the overall fake decision for the entire input.


The utterance begins with a real segment and transitions to a fake one. As we analyzed in the introduction, the FLD model heavily relies on detecting splicing artifacts, a fragile strategy that can fail with more sophisticated attacks.

In contrast, the SAL model delivers a stable and confident prediction, with its score remaining decisively below the decision threshold across the entire spoofed segment. The corresponding Grad-CAM heatmap provides compelling evidence for its advanced reasoning: rather than focusing on the boundary, SAL displays high, sustained activations deep within the spoofed content itself. This demonstrates that our proposed model has successfully learned to identify the intrinsic properties inherent to the synthetic signal.

\begin{figure}[t]
    \centering
    \begin{subfigure}[b]{0.42\linewidth}
        \centering
        \includegraphics[width=\linewidth]{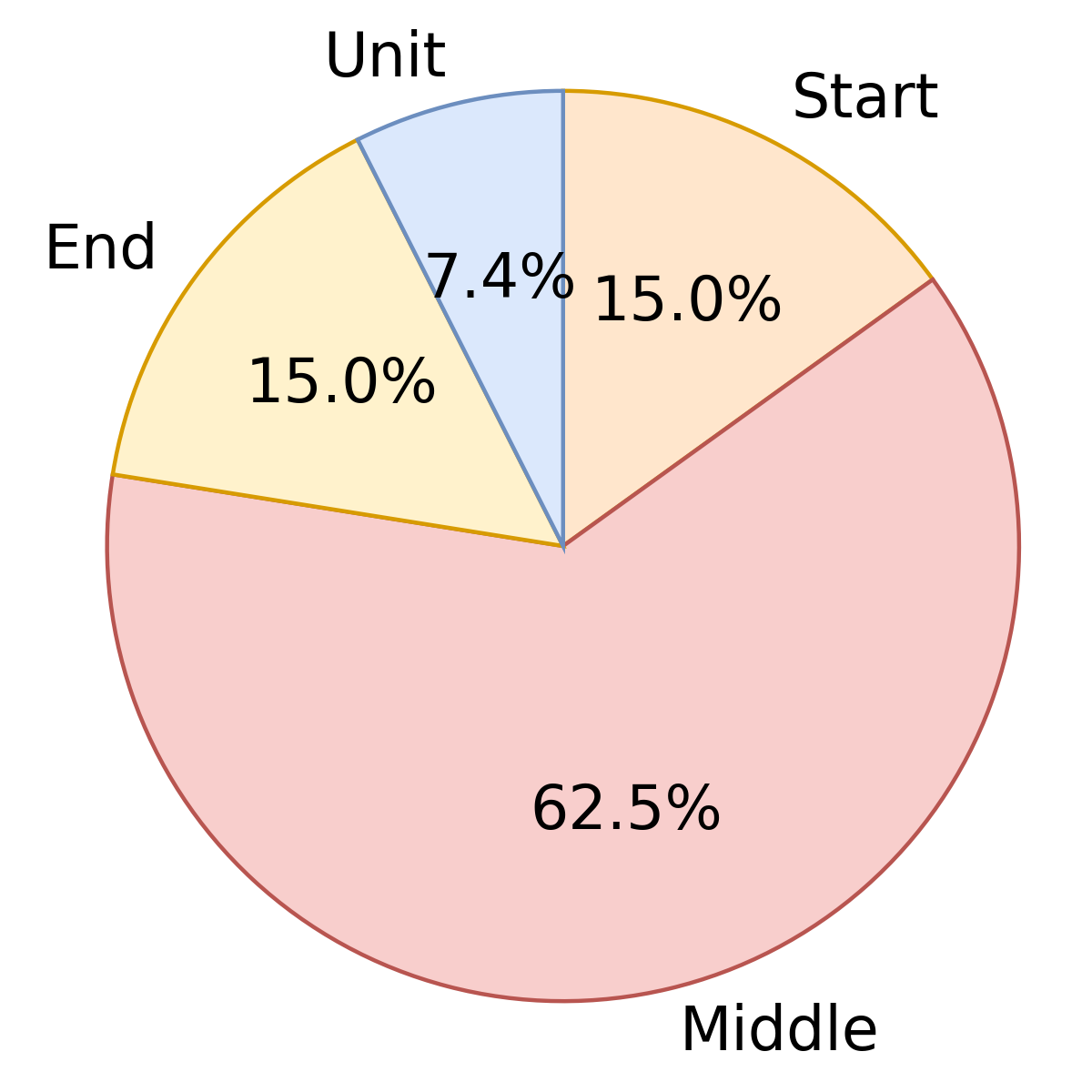}
        \caption{Distribution.}
        \label{fig:spoof_distribution}
    \end{subfigure}
    \hfill 
    \begin{subfigure}[b]{0.55\linewidth}
        \centering
        \includegraphics[width=\linewidth]{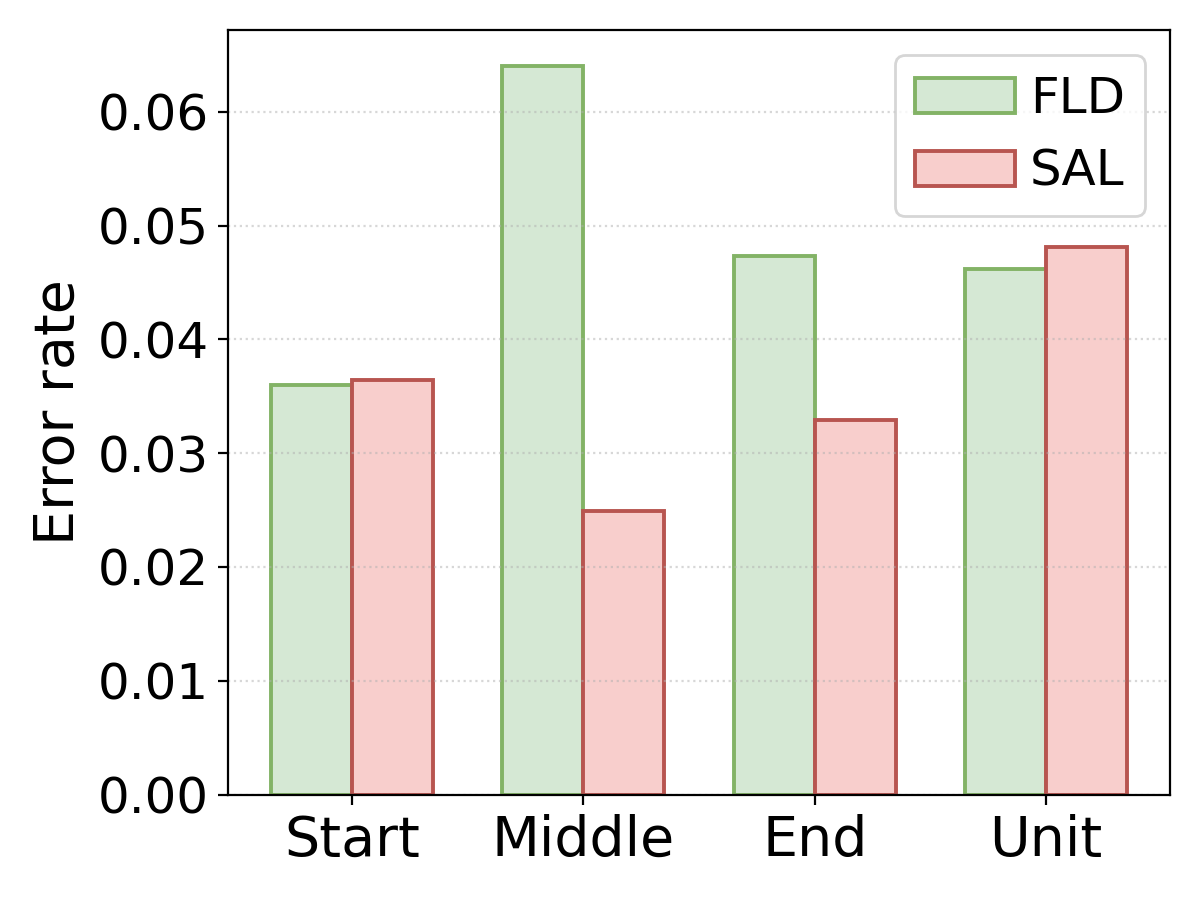}
        \caption{Error rate comparison.}
        \label{fig:error_comparison}
    \end{subfigure}
    \caption{Visualization of model performance on different positions. (a) shows that middle-segments are the most common case. (b) demonstrates that our SAL model significantly reduces errors on this specific category compared to the FLD model.}
    \label{fig:visualization}
    \vspace{-10pt}
\end{figure}


The phenomenon revealed in this case study is not an isolated incident but a systemic improvement. To quantitatively validate this advantage from a holistic perspective, we further analyzed the models' performance on segments at different utterance positions, as shown in Fig.~\ref{fig:visualization}.

The pie chart in Fig.~\ref{fig:spoof_distribution} shows the distribution of positions. We can find that middle segments are the most common, accounting for the majority of cases (62.5\%). The bar chart in Fig.~\ref{fig:error_comparison} compares the error rates of our final SAL model against the FLD baseline for each location. It is clear the FDL system struggles significantly with middle segments. In stark contrast, our SAL model achieves its greatest performance improvement precisely in this category. 

Together, these analyses form a complete chain of evidence. The progression from the specific case of attention visualization to the overall view of statistical error analysis jointly proves that our segment-aware approach effectively overcomes shortcut learning to accurately detect the most common and challenging types of partial spoofing attacks.
\section{Conclusion}
\label{sec:concl}

In this work, we address the limitations of transition-centric approaches in partial deepfake localization by introducing Segment-Aware Learning. By integrating Segment Positional Labeling and Cross-Segment Mixing, SAL learns the intrinsic characteristics of entire audio segments. Our approach achieves strong and consistent performance across multiple datasets in both in-domain and cross-domain scenarios. Analysis through Grad-CAM visualizations and error breakdowns further confirms that SAL improves detection accuracy in non-boundary regions and mitigates shortcut learning based on transition artifacts. These results highlight the effectiveness of segment-aware learning over boundary-focused strategies in detecting partial audio deepfakes.

\vfill\pagebreak

\bibliographystyle{IEEEbib}
\bibliography{refs}

\end{document}